\documentclass[twocolumn,showpacs,preprintnumbers,amsmath,amssymb]{revtex4}
\usepackage{epsfig,psfrag,graphicx,hyperref}
\begin{document}

\title{Stimulated Raman and Brillouin backscattering of collimated beams carrying orbital angular momentum}

\author{J.T. Mendon\c{c}a}

\email{titomend@ist.utl.pt}

\affiliation{IPFN, Instituto Superior T\'{e}cnico, Av. Rovisco Pais 1, 1049-001 Lisboa, Portugal}

\author{B. Thid\'e}

\altaffiliation[Also at ]{LOIS Space Centre, V\"axj\"o University,
  SE-351\,95 V\"axj\"o, Sweden}
\affiliation{Swedish Institute of Space Physics, {\AA}ngstr\"{o}m
  Laboratory, P.\,O.~Box~537, SE-751\,21, Uppsala, Sweden}

\author{H. Then}

\affiliation{Institute of Physics, Carl von Ossietzky University, 26111 Oldenburg, Germany}

\begin{abstract}

We study theoretically the exchange of angular momentum between electromagnetic and electrostatic waves in a plasma, due to the stimulated Raman and Brillouin backscattering processes.  Angular momentum states for plasmon and phonon fields are introduced for the first time. We demonstrate that these states can be excited by nonlinear wave mixing, associated with the scattering processes. This could be relevant for plasma diagnostics, both in laboratory and in space. Nonlinearly coupled paraxial equations and instability growth rates are derived.

\end{abstract}

\maketitle

%\pacsnumbers{}

%\section{Introduction}

It is well known that the angular momentum of electromagnetic radiation contains two distinct parts, one associated with its polarization state, or photon spin, the other being the external or orbital photon angular momentum (OAM). With the advent of laser beams, 
an increasing interest is being given to the study of photon OAM, and various optical experimental configurations have been considered \cite{allen,harris,padgett,leach}. It is now well understood that collimated electromagnetic beams, such as laser or radio wave beams,  can be described by Laguerre-Gaussian functions, which provide a natural orthonormal basis for a generic beam representation. Utilization of photon OAM states in the low frequency ($ \leq \, \operatorname{GHz}$) radio wave domain was also recently proposed in Ref.~\onlinecite{thide}, as a new method for studying and characterizing radio sources in astrophysics.

The possibility of remote study of space plasma vorticity by measuring the OAM of radio beams interacting with vortical plasmas was pointed out in Ref.~\onlinecite{thide2},  and a more detailed theoretical analysis was given recently by studying 
the electromagnetic wave scattering from the plasma medium, with the associated OAM exchanges between  the plasma and probing photon beams \cite{mendBo1}.  A more speculative work was also recently published where the strong similarities between photon and neutrino dispersion relations were explored, and OAM states of neutrino beams interacting with dense plasmas were considered \cite{mendBo2}.

Here we consider the important problem of stimulated Raman and Brillouin backscattering of collimated electromagnetic beams with finite OAM in a plasma. This also leads us to consider, to our knowledge for the first time, the possible existence of plasmon and phonon states with finite orbital angular momentum. Raman and Brillouin scattering instabilities are well known in the context of laser fusion \cite{kruerbook}, as possible sources of anomalous plasma reflectivity. Raman backscattering is now recognized as a dominant process for ultra-intense laser plasma interactions, in the context of inertial fusion research \cite{yin}. In all these studies, angular momentum in general, and photon OAM in particular, have been systematically ignored.  On the other hand, there seems to be experimental evidence of OAM dependence in Brillouin scattering of radio waves in the ionosphere \cite{norin}, which awaits for a deeper theoretical understanding. 

In contrast with the traditional theoretical approach \cite{kruerbook}, we consider the case of an electromagnetic pump beam with finite transverse dimension, and arbitrary OAM states are considered. Our formalism is quite simple but general, and includes the nonlinear coupling between incident and backscattered waves with electron plasma waves (for the Raman instability) or ion acoustic waves (for the Brillouin scattering). Nonlinear paraxial equations for the electromagnetic wave modes will be coupled to paraxial equations for the electrostatic wave modes. These later equations will allow us to introduce the angular momentum states for both plasmons and phonons. Notice that the spin effects are absent for the electrostatic oscillations, which means that (in contrast with the photon case) their OAM coincides with their total angular momentum.

In the following, instability growth rates for generic OAM states of the incident or collimated pump beams will be determined, for an infinite and homogeneous plasma. An important point to notice is that, even in the infinite plasma, our theoretical model will predict a well localized backscattering instability, taking place at the focal region of the incident beam. 

%\section{Basic equations}
In isotropic and homogeneous plasmas, where the ions are assumed immobile, transverse and longitudinal wave coupling can be described by the following wave equations
\begin{equation}
\left( \frac{\partial^2}{\partial t^2} - c^2 \nabla^2 + \omega_{pe}^2 \right) \vec{A} = - \frac{e^2}{\epsilon_0 m} \tilde{n} \vec{A}
\label{eq:1} \end{equation}
and
\begin{equation}
\left( \frac{\partial^2}{\partial t^2}  - S_e^2 \nabla^2 + \omega_{pe}^2 \right) \tilde{n} = \frac{n_0 e^2}{2 m^2} \nabla^2 A^2
\label{eq:2} \end{equation}
where $\vec{A} \equiv \vec{A} (\vec{r}, t)$ is the vector potential describing the transverse electromagnetic waves, $\tilde{n}$ the electron density perturbations associated with the electrostatic waves, $n_0$ the equilibrium electron density, $\omega_{pe} = (e^2 n_0 / \epsilon_0 m)^{1/2}$ is the electron plasma frequency, $-e$ and $m$ are the electron charge and mass, $\epsilon_0$ the vacuum permittivity, $c$ the speed of light, and $S_e = \sqrt{3 T_e / m}$ the electron thermal velocity, for a plasma with electron temperature $T_e$.
 
We now assume wave propagation along the $z$-direction, by considering wave solutions of the form
$\vec{A} = \sum_{j = 1, 2} \vec{A}_j \exp (i k_j z - i \omega_j t) + c.c.$, where $\omega_j$ and $k_j$ are the frequencies and wavenumbers of the two electromagnetic wave modes (the incident and the scattered one). Similarly, we can use for the electrostatic oscillations
$\tilde{n} = \tilde{n}_0 \exp (i k' z - i \omega' t) + c.c.$. We assume that the wave amplitudes $\vec{A}_j$ and $\tilde{n}_0$ are slowly varying on space and time scales much longer than the respective wavelengths and periods. Using such an assumption, we can reduce the wave equations (\ref{eq:1}) and (\ref{eq:2}) to three coupled equations of the form
\begin{eqnarray} \label{eq:3}
&D_1 \vec{A}_1 = \omega_{pe}^2 ( \tilde{n}_0 / n_0) \vec{A}_2 \; , \quad 
D_2 \vec{A}_2 = \omega_{pe}^2 ( \tilde{n}_0^* / n_0) \vec{A}_1 
\\  \nonumber
& D'  \tilde{n}_0 = n_0 ( e^2 k^{'2} / 2  m^2 ) \left( \vec{A}_1 \cdot \vec{A}_2^* \right) 
\end{eqnarray}
where we have used the following operators
\begin{eqnarray} \label{eq:4}
&D_j = c^2 \left( \nabla_\perp^2 + 2 i k_j \partial / \partial z \right) + 2 i \omega_j \partial / \partial t \; ,
\\ \nonumber
&D' = S_e^2 \left( \nabla_\perp^2 + 2 i k' \partial / \partial z \right) + 2 i \omega' \partial / \partial t 
\end{eqnarray}
for $j =1, 2$,  we have assumed the energy and momentum conservation relations $\omega_1 = \omega_2 + \omega'$ and $k_1 = k_2 + k'$, and that the linear dispersion relations are satisfied, for the transverse and electrostatic modes,
$k_j^2 c^2 = (\omega_j^2 - \omega_{pe}^2)$, and  $k^{'2} S_e^2 = (\omega^{'2} - \omega_{pe}^2)$, respectively. We should keep in mind that $\omega' \simeq \omega_{pe}$, and consequently the incident wave frequency is larger than twice this value, $\omega_1 \geq 2\omega_{pe}$. It is known that, in order to make all these conditions compatible with each other, we have to assume that the two electromagnetic wave modes propagate in opposite directions, with $k_1 > 0$ for the incident wave, and $k_2 = - | k_2 | < 0$ for the backscattered wave, with $k' > 0$ for the electrostatic wave.
\begin{figure}
		 \includegraphics[angle=0,scale=0.75]{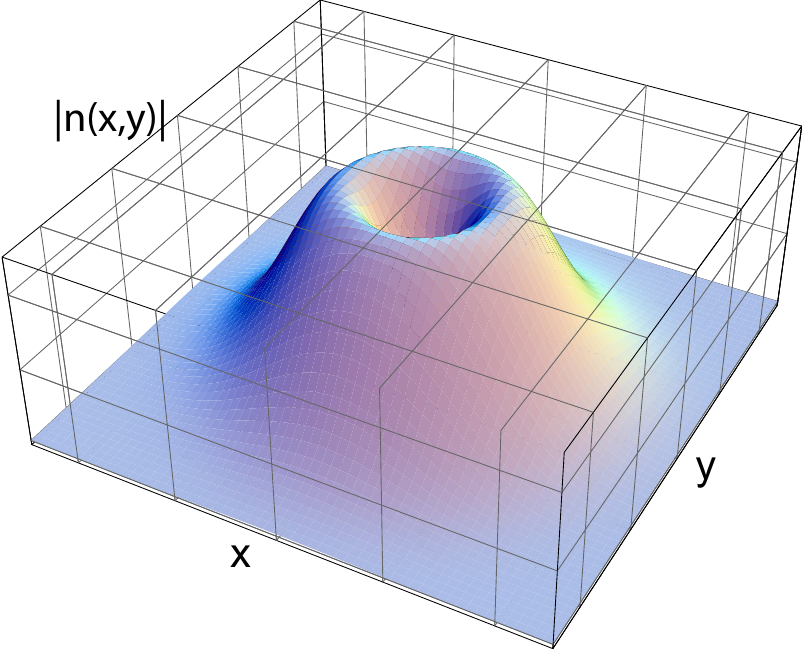}
	\caption{\label{fig1}  {\sl Amplitude of the electron density fluctuations associated with plasmon states of finite (orbital) angular momentum, for $l' = 1$ and $p' = 0$, in arbitrary units.}}
\end{figure}
%\section{Photon and plasmon OAM states}
Let us first discuss equations (\ref{eq:3}) in the linear approximation, where the coupling terms on the r.h.s. are neglected. The temporal dependence of the amplitudes $\vec{A}_j$ and $\tilde{n}_0$ will vanish, and these linear equations will reduce to pure paraxial equations, taking the form
\begin{equation}
\left( \nabla_\perp^2 + 2 i k_j \frac{\partial}{\partial z} \right) \vec{A}_j = 0 \; , \quad
\left( \nabla_\perp^2 + 2 i k' \frac{\partial}{\partial z} \right) \tilde{n}_0 = 0
\label{eq:6} \end{equation}
Using cylindrical coordinates $\vec{r} \equiv (r, \varphi, z)$, the paraxial wave solutions for the first of these equations can be written as linear combinations of modes 
$\vec{A}_j (\vec{r}) = \vec{A}_{p_j, l_j} (z) F_{p_j, l_j} (r, z) e^{i l_j \varphi}$, where $F_{p_j, l_j} ( r, z)$ are Laguerre-Gaussian functions, with integers $p_j, l_j$  representing the radial and azimuthal (quantum) numbers, as defined by
\begin{equation}
F_{p_j, l_j} ( r, z) = \frac{1}{2 \sqrt{\pi}} \left[ \frac{(l_j + p_j)!}{p_j !} \right]^{1/2} x_j^{|l_j|} L_{p_j}^{|l_j|} ( x_j) 
\label{eq:6b} \end{equation}
where $w_j \equiv w_j (z)$ are the beam waists of the two electromagnetic wave modes, $j = 1, 2$, $x_j = ( r / w_j )^2$, and $L_p^l (x)$ are associated Laguerre polynomials.
Due to the well known orthogonality relations of the Laguerre-Gaussian modes, we can write
$\int_0^\infty r  dr F_{p_1,l_1} (r, z) F^*_{p_2, l_2} (r, z) \int_0^{2\pi} d \varphi e^{ i (l_1 - l_2) \varphi} = \delta_{p_1, p_2} \delta_{l_1, l_2}$,
where the deltas represent Kronecker symbols.
It is also known that these solutions of the vector potential $\vec{A}_j$ correspond to well defined photon orbital angular momentum states, characterized by the azimuthal quantum numbers $l_j$. Similarly, we can say that the $\tilde{n}_0$ solutions correspond to a superposition of plasmon angular momentum states, characterized by the radial and azimuthal quantum numbers $p', l'$,  of the form $F_{p', l'} (r, z) e^{i l' \varphi}$. The beam waist $w' \equiv w' (z)$ of the corresponding plasmon modes is considered here to be of the same order as the beam waist of the dominant electromagnetic wave mode, $w' \simeq w_1$. Notice however that the electrostatic waves carry no intrinsic angular momentum, because in contrast to the transverse photons, plasmons have spin zero. This means that the plasmon angular momentum states coincide with their total angular momentum states. 
The electric field perturbations associated with these plasmon states will remain purely electrostatic, if they satisfy $\nabla \times \vec{E} = 0$. For solutions of the form $\vec{E} (\vec{r}, t) = \vec{E} (r) \exp (i l' \varphi + i k' z - i \omega' t)$, this will imply the existence of angular and radial field components such that $E_\varphi = l' E_z / k«r$, and $E_r = - i d E_z / k' d r$, where $E_z$ is the axial field component. The same will be true for the electron mean velocity $\vec{v} = - i (\epsilon_0 \omega' / e n_0) \vec{E}$. In the present calculations we only need to use the explicit form for the electron density fluctuations $\tilde{n}_0$, as described by the paraxial equation (\ref{eq:6}). The corresponding solution can be described as a superposition of orthogonal LG modes 
\begin{equation}
\tilde{n}_0 (\vec{r}, t) =  \tilde{n}_{p', l'} (z) F_{p', l'} (r, z) e^{i l' \varphi} e^{i k' z - i \omega' t} \vec{e}_z + c.c.
\label{eq:8b} \end{equation}
These plasmon modes can then take quite unusual shapes, as illustrated in figure 1.
%\section{Photon-plasmon angular momentum coupling}
Going back to the coupled nonlinear equations (3), we can see that they describe an exchange of energy, linear momentum, as well as orbital angular momentum, between the two electromagnetic waves and the longitudinal waves. Replacing the linear solutions (with amplitudes now depending on time) in these equations, and integrating over the radial variable $r$, we arrive at the following nonlinear coupled equations 
\begin{eqnarray} \label{eq:9}
&\partial \vec{a}_1 / \partial t = - i C_1 n' \vec{a}_2  \quad , \quad
\partial \vec{a}_2^* / \partial t = i C_2 n' \vec{a}_1^*
\\ \nonumber
&\partial n' / \partial t = - i C' ( \vec{a}_1 \cdot \vec{a}_2^*)
\end{eqnarray}
with the nonlinear coupling coefficients
\begin{equation}
C_j = \pi \frac{e^2 R}{\epsilon_0 \omega_j m} \; , \quad C' = \frac{\pi}{2} \frac{n_0 e^2}{m^2} \frac{k^{'2}}{\omega'} R
\label{eq:9b} \end{equation}
We have also used the simplified notation $\vec{a}_j = \vec{A}_{p_j,l_j} (z , t)$ and $n' = \tilde{n}_{p', l'} (z, t)$, and introduced the quantity
$R \equiv R (z) = \int_0^\infty F_{p_1,l_1} F_{p_2, l_2} F_{p', l'} r dr $. 
In order to understand the physical meaning of these coupled equations for the Laguerre-Gaussian modes, let us consider the important case of stimulated Raman scattering, by considering an intense incident wave with amplitude $a_1$, and by using the parametric approximation $\partial a_1 / \partial t \simeq 0$. We also assume maximum coupling conditions, corresponding to parallel polarization $\vec{a}_1 \parallel \vec{a}_2$. From the two remaining equations, we can then easily derive
\begin{equation}
\frac{\partial^2 n'}{\partial t^2} = \gamma^2 n' \; , \quad  \gamma = \omega_{pe} \frac{\pi e k'}{m \sqrt{2 \omega' \omega_2}} | a_1| R
\label{eq:10a} \end{equation}
This leads to the following unstable solutions $n' (z, t) = n' (z, 0) \exp (\gamma t)$, and $a_2 (z, t) = a_2 (z, 0) \exp (\gamma t)$, with growth rate $\gamma$. They show that the incident wave with amplitude $a_1$ can excite electron plasma waves, or plasmons, with different angular momentum states. In the case of plasmons carrying no angular momentum, the backscattered wave will be in a state $l_2 = - l_1$. The observation of a 
backscattered signal with $l_2 \neq - l_1$ will therefore reveal the existence of plasmons with non-zero angular momentum states propagating in the medium. Raman backscattering can then be used as a powerful diagnostic method to detect internal plasma vorticity. 
\begin{figure}
		 \includegraphics[angle=0,scale=1]{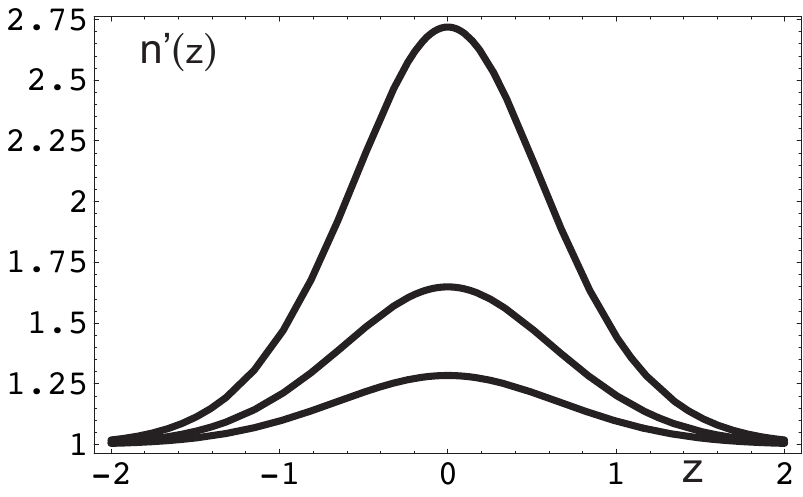}
	\caption{\label{fig2} {\sl Normalized amplitude of the plasmon mode excited across the laser focal region, for three different times.}}
\end{figure}
Let us now consider the interesting situation where well defined plasmon angular momentum states can be excited from the outside. This can be done by using two counter-propagating electromagnetic waves. Going back to equations (\ref{eq:9}), we can determine the amplitude of the excited plasmon state, as a function of the initial backscattered wave signal $a_2 (z, 0)$. Again, we assume $|a_1 | \gg |a_2 |$ and use the parametric approximation. Equations (\ref{eq:9}), for parallel polarization, can then be reduced to
\begin{equation}
\frac{\partial a_2^*}{\partial t} = i C_2 n' a_1^* e^{- i \Delta \omega t} \; , \quad \frac{\partial n'}{\partial t} = - i C' a_1 a_2^* e^{ i \Delta \omega t}
\label{eq:11} \end{equation}
Here we have introduced a finite frequency mismatch $\Delta \omega = \omega_1 - \omega_2 - \omega'$, because the two electromagnetic wave modes imposed from the outside are not necessarily in the optimum matching conditions. We can then derive 
\begin{equation}
\frac{\partial^2 n'}{\partial t^2} =  \gamma^2 n' + i \Delta \omega \; e^{ i \Delta \omega t} \frac{\partial n'}{\partial t}
\label{eq:10b} \end{equation}
again with $\gamma^2 = C' C_2 |a_1|^2 R^2$. For initial conditions such that $n' (z, 0) = 0$ and $a_2 (z, 0)$ is arbitrary (but still obeying the linear paraxial solution along $z$), we can write the solution of these equations in the form
\begin{equation}
a_2 (z, t) = a_2 (z, 0) \cosh (g t) e^{ i \Delta \omega t / 2}
\label{eq:12} \end{equation}
with $g = [ \gamma^2 - (\Delta \omega / 2)^2 ]^{1/2}$, and 
\begin{equation}
 n' (z, t) = - i \frac{C' a_1}{g} a_2 (z, 0) \sinh (g t) e^{ i \Delta \omega t / 2}
\label{eq:12b} \end{equation}
These solutions describe the growth of the backscattered signal with orbital angular momentum state $l_2$, and the excitation of a plasmon angular momentum state characterized by the azimuthal number $l' = l_1 - l_2$. 
Notice that the above growth rates depend on the axial position $z$. This means that, very soon, the axial profile of both the excited backscattered and electrostatic modes will deviate from its linear solution, as illustrated in figure 2. However, the radial beam profile will remain unchanged.  

Stimulated Brillouin scattering with orbital angular momentum can be treated in a similar ground. In this case, the nonlinear plasmon equation (\ref{eq:2}) is replaced by the nonlinear equation
\begin{equation}
\left( \frac{\partial^2}{\partial t^2} - v_s^2 \nabla^2  \right) \tilde{n} = \frac{Z n_0 e^2}{m M} \nabla^2 A^2
\label{eq:13} \end{equation}
where $\tilde{n}$ now represents the electron density oscillations associated with the ion acoustic waves, $v_s = \sqrt{Z T_e / M}$ is the ion acoustic velocity, and $Z e$ and $M$ are the ion charge and mass. Following the same procedure, and assuming that $\tilde{n} = \tilde{n}_0 \exp (i k' z - i \omega' t) + c.c.$, we can arrive at the nonlinear paraxial equation for the ion acoustic wave amplitude
\begin{equation}
D' \tilde{n}_0 = \frac{Z n_0 e^2 k^{'2}}{m M} \left( \vec{A}_1 \cdot \vec{A}_2^* \right) 
\label{eq:14} \end{equation}
where the differential operator is the same as before, but with the electron thermal velocity $S_e$ replaced by the ion acoustic velocity $v_s$, and with the parallel wavenumber $k'$ satisfying now the ion acoustic dispersion relation $k' v_s = \omega'$. In the linear approximation, we recover the electrostatic paraxial equation, similar to the second of equations (\ref{eq:6}). This shows that angular momentum states for ion acoustic waves, formally identical to those of plasmons, can be defined such that $\tilde{n}_0 (\vec{r} ) = n' F_{p', l'} (r, z) e^{i l' \varphi}$, with $n' = \tilde{n}_{p', l'} (z, t)$. Inserting these linear solutions,  in the nonlinear paraxial equation (\ref{eq:13}), we get for the time dependent ion acoustic wave amplitude
\begin{equation}
\frac{\partial n'}{\partial t} = - i C_B a_1 a_2^* \; , \quad C_B = \pi \frac{Z n_0 e^2}{m M} \frac{k'}{ v_s} R
\label{eq:15} \end{equation}
This leads to stimulated Brillouin backscattering solutions, with a growth rate $\gamma_s$, determined by $\gamma_s^2 = C_B C_2 | a_1 |^2$. All the qualitative features discussed above for the Raman backscattering case can be repeated here. In particular, the use of electromagnetic wave scattering as a diagnostic probe of internal plasma vorticity, and the excitation of well defined angular momentum states of the phonon spectrum by two counter-propagating electromagnetic waves.

%\section{Conclusions}

In this work we have considered stimulated Raman and Brillouin backscattering of electromagnetic waves in a plasma with photon orbital angular momentum. We have also introduced, to our knowledge for the first time, the related concepts of plasmon and phonon angular momentum states. The field modes associated with these electrostatic wave modes are determined by solutions of appropriate paraxial equations, similar to those describing collimated electromagnetic wave beams near the focal region.  In the case of these electrostatic quasi-particles, the orbital angular momentum coincides with the total angular momentum, because, in contrast to photons, plasmons and phonons have no spin.

We have discussed the nonlinear coupling between incident and backscattered radiation in uniform plasmas, and derived the corresponding growth rates, in the parametric approximation. These results generalize the well known results of stimulated Raman and Brillouin scattering, by including the finite size and radial profile of the wave modes, corresponding to the various orbital angular momentum states. We have shown that an additional selection rule for nonlinear wave interaction, associated with the conservation of angular momentum, was added to the usual energy and linear momentum selection rules. In particular, we have shown that, by using two counter-propagating electromagnetic waves with well defined orbital angular momentum, we can excite specific states of non-zero plasmon and phonon angular momentum. Experimental verification of our theoretical model could lead to a novel result on basic plasma physics, to be considered in the future.   

In this work we have only considered three wave coupling processes. But the present formalism can easily be extended to four wave coupling, which could be particularly interesting for the case of Brillouin scattering. Notice that, although the present model only applies to uniform plasmas, the unstable region where stimulated scattering takes place is highly localized, first because of the finite transverse width of the electromagnetic beams, second because of the enhancement of the instability growth rate in the axial direction, over a distance eventually much shorter than the Rayleigh length. 

The Swedish author (B. T.) gratefully acknowledges the financial support from the Swedish Research Council (VR).


\begin{thebibliography}{10}

\bibitem{allen}
L. Allen, M.W. Beijersbergen, R.J.C. Spreeuw, and J.P. Woerdman, {\sl Phys. Rev. A}, {\bf 45}, 8185 (1992).

\bibitem{harris}
M. Harris, C.A. Hill, P.R. Tapster, and J.M. Vaughan, {\sl Phys. Rev. A}, {\bf 49}, 3119 (1996). 

\bibitem{padgett}
M.J. Padgett, J. Arlt, N.B. Simpson, and L. Allen, {\sl Am. J. Phys.}, {\bf 64}, 77 (1996).

\bibitem{leach}
J. Leach, M.J. Padgett, S.M. Barnett, S. Franke-Arnold, and J. Courtial, {\sl Phys. Rev. Lett.}, {\bf 88}, 257901 (2002).

\bibitem{thide}
B. Thid\'e, H. Then, J. Sj\"oholm, K. Palmer, J. Bergman, T.D. Carozzi, Ya. N. Istomin, N.H. Ibragimov, and R. Khamitova, {\sl Phys. Rev. Lett.}, {\bf 99}, 087701 (2007).

\bibitem{thide2}
B. Thid\'e, {\sl Plasma Phys. Control. Fusion}, {\bf 49} (128), B103-B107 (2007).

\bibitem{mendBo1}
J.T. Mendon\c{c}a, B. Thid\'e et al., {\sl Phys. Rev. Lett.}, submitted (2008).

\bibitem{mendBo2}
J.T. Mendon\c{c}a and B. Thid\'e et al., {\sl Europhys. Lett.}, {\bf 84}, 41001 (2008).

\bibitem{kruerbook}
W. Kruer, {\sl The Physics of Laser Plasma Interactions}, Addison-Wesley Publishing Company, Redwood City CA (1973).

\bibitem{yin}
  L. Yin et al., {\sl Phys. Plasmas}, {\bf 15}, 013109 (2008). 

\bibitem{norin}
L. Norin, T. B. Leyser, E. Nordblad, B. Thid\'{e}, and M. McCarrick,
{\sl Phys. Rev. Lett.}, {\bf 106}, 065003 (2009).

\end{thebibliography}
\end{document}